\documentclass[epj]{svjour}
\usepackage{graphics}
\begin{document}
\title{The Quark--Hadron Transition in the Early Universe}

\author{Giandomenico Sassi\inst{1} \& Silvio A. Bonometto\inst{1,2}}
\institute{Physics Department G. Occhialini, Universit\`a degli Studi di
Milano--Bicocca, Piazza della Scienza 3, I20126 Milano (Italy)
\and I.N.F.N., Sez.~Milano--Bicocca,
Piazza della Scienza 3, I20126 Milano (Italy)}


\date{Received: ~~~~~~~~~~~~~ / Revised version: ~~~~~~~~~~~~~}

\abstract{
We use recent lattice QCD outputs to work out the expansion law of the
Universe during the cosmological quark--hadron transition. To do so, a
suitable technique to exploit both pressure and energy density data,
with the related error bars, is introduced. We also implement suitable
techniques to relate the $T$ range where lattice outputs are available
with lower and higher $T$'s, for which we test suitable expressions.
We finally compare the cosmological behavior found using lattice data
with the one obtainable in the case the transition were first order,
although not so far from the crossover transition we studied.
Differences are small to be tested with cosmological data, but the
coming of the era of precision cosmology might open a channel to
inspect the QCD transition through them.
\PACS{
{12.38.Gc}{lattice QCD calculations} \and
{90.80-k}{cosmology}
}}

\maketitle

\section{Introduction}
In the very early Universe, strongly interacting matter (SIM) was a
quark--hadron plasma. Then cosmic expansion lowered the temperature
$T$ to values ${\it O}$$(100$--$200\, $MeV$)$ and, for a short period,
SIM turned into a hadron gas made of $\pi$ and mesonic resonances,
plus rare nucleons carrying the cosmic baryon number $B$. We shall
refer to this process as Cosmic Quark--Hadron transition (CQHt). Soon
after, when $T$ shifts below $ m_\pi$ (pion mass), the only residual
SIM shall be made of the baryons needed to carry $B$.

CQHt might leave an imprint on today's cosmic observables. This idea
was widely explored in the mid-Eigh\-ties, when the option that CQHt had
been a first order phase transition was considered likely. In that
time lattice QCD outputs were already tentatively used to work out the
cosmic expansion rate and the time dependence of the scale factor and
several authors also obtained analytical determination of the time
dependence of the scale factor, $a(t)$, during a first order phase
transition, occurring close to the critical temperature $T_c$
\cite{bonlat}. One of the main ingredients to find a right solution of
these problems was taking into account the presence of the
lepton--photon component in the CQHt epoch.

Lattice outputs were considered to this very aim a few times and even
quite recently (see, {\it e.g.}, \cite{latexp}). Often, a technique to
exploit them was re--invented and a few errors occurred recursively,
{\it e.g.} the neglect of the lepton--photon component.

A topic of this work will then be a general outline of the technique
and the problems to exploit lattice data to cosmological aims.

A critical issue, concerning CQHt, was however risen by Witten
\cite{witten} in the mid-Eighties, and this made then CQHt a hot
subject. He showed that, in the case of a first order transition, $B$
would tend to remain in the plasma. This led him to suggest that {\it
quark nuggets} could still exist in today's Universe, constituting the
cosmic Cold Dark Matter (CDM) component. The apparent similarity
between the present densities of the baryon and CDM components could
then find a satisfactory explanation. This idea has been widely
debated and the interest for it never completely faded (see, {\it
e.g.}, \cite{lugones}).

Other researchers, although keeping to the idea that $B$ kept in the
plasma, until it existed, suggested that the transition would however
reach its completion, leading however to an inhomogeneous $B$
distribution. Neutron diffusion would then rapidly smooth out their
inhomogeneities. Proton diffusion, instead, requires a coherent
electron flow and the electron Thomson cross section is so large that
proton inhomogeneities may last down to a temperature $\sim 50\, $keV,
being therefore able to affect Big--Bang Nucelosynthesis (BBN)
\cite{BBN1}. This idea triggered a large deal of works \cite{BBN2}, of
whom we provide a (partial) list up to 1990.

As a matter of fact, a first order phase transition occurs in lattice
QCD without dynamical quarks \cite{quenched}. For dynamical quarks
with vanishing mass, the transition is also first order. These outputs
were given a great importance until physical mass quarks could not be
treated on the lattice. Then, if {\it up} and {\it down} quarks are
light and the {\it strange} quark is massive, according to
phaenomenological results, lattice outputs unequivocously yield no
confinement phase transition, but a smooth crossover.

Some doubts can still exist on the nature of the transition leading to
the breaking of chiral symmetry. Recent speculation related this
transitions to the origin of cosmic magnetic fields \cite{devega},
although a problem exists on their scale, which keeps related to the
horizon scale at $t \simeq 10^{-5}\, $sec..

If CQHt is not a first order phase transition, its relevance for
cosmology is no longer so great. It must also be outlined that the
recent analysis of CMB, BAO and SNIa data \cite{komatsu}, within the
frame of a $\Lambda$CDM cosmology, has determined a reduced baryon
density parameter $\omega_b = 0.0227 \pm 0.0006$, in fair agreement
with what is predicted by homogeneous BBN, in accordance with the
observational values of $^2H/H$ and $^4He/H$ ratios. 

Some discrepancies however exist with observational $^3He$ and $^7Li$
abundances. They are likely to arise from marginal errors either in
observations or in specific nuclear reaction rates. For instance,
slight shifts in the rates of the $^7Be(d,p)2^4He$ or
$^3He(\alpha,\gamma) ^7Be$ reactions \cite{coc} could put a remedy to
the $^7Li$ discrepancy.

The possibility that the observational values of $^2H/H$ and $^4He/H$
ratios are an indication of inhomogeneous BBN, or derive from other
peculiarities as particle decays during of after BBN, is still open,
but is really second choice \cite{olive}.

On the contrary, in the late Eighties the situation was not so
strongly constrained; open cosmologies were then considered a valid
option, while $\Lambda$CDM was mostly considered just as a
counter--example in n--body simulations. The underlying idea, in the
study of inhomogneous BBN was that is could relax the constraints on
$\omega_b$, making it compatible with a present overall density
parameter $\Omega_0 \sim 0.2$, so avoiding the need of non--baryonic
DM.

The present situation is however such that, should fresh data reopen
the option of first order CQHt, we should then take care that its
features do not produce too strong proton inhomogeneities, radically
perturbing the theory--observation agreement, as far as BBN is
concerned. Only in this sense, perhaps, BBN remains a constraint to
CQHt.

Let us then also remind that, although lattice QCD indicates that CQHt
was crossover, this is still the outcome of a theoretical elaboration.
No experimental confirm can be soon available and the most direct
pattern to explore this physics is likely to be just cosmology. With
the arrival of the era of precision cosmology, however, this might
become a realistic option.

Besides of trying to provide a general setup on the use of lattice
outputs in cosmology, this paper therefore aims at evaluating the
impact that different CQHt patterns could have on cosmic observables.
We shall then also consider the possibility that CQHt is a first order
phase transition, keeping however quite close to recent lattice
outputs, and will make a comparison between the cosmic evolutions in
the two cases.

A further point we need to explore is the connection between the very
high $T$ regime, where asymptotic freedom is approached, and the
$T$--range where lattice outputs are available. We shall show that an
analytical expression, which can be also seen as a generalization of
the historical Bag Model \cite{13}, is able to fit high--$T$ lattice
outputs, connecting them with the temperature range where the plasma
behaves as an ideal gas. The procedure to provide this expression and
to fit it to data is one of the outputs of this work.

Finally, we shall also need an expression to describe SIM in the very
low $T$ regime, when it has turned into a hadron gas. Although SIM is
then just a minor component of the Universe, its impact on the cosmic
expansion is not yet fully negligible. In this paper we shall describe
SIM, in the form of a hadron gas, by using a state equation inspired
to the Hagedorn \cite{hagedorn} model, whose parameters will be
selected in order to connect it with the state equation resulting from
lattice outputs.

Back in the Eighties, a large deal of work concerned also bubble
nucleation, expansion, and coalescence, during CQHt (for a review see,
{\it e.g.}, \cite{bonpan}). In the absence of a phase transition they
are hardly relevant.

Recent outputs on SIM bulk viscosity during ion collisions
\cite{tawfik}, however, could lead to conjecture some more intricate
situation. Viscosity is an indication that particle reactions proceed
too slowly to allow the system to settle in the equilibrium
configurations of an ideal fluid. A tentative data interpretation
could then be that reactions putting together 3 quarks have a low rate
and, perhaps, require an intermediate di--quark state. In this case
their rate would be controlled by the average di--quark concentration
in high--$B$ quark gluon plasma.

Should this interpreation of bulk viscosity be correct, we expect
viscosity to be negligible in the low--$B$ cosmological context.

One might however wonder whether the low rate for $3q \to B$ reactions
could however mean that $B$ keeps being carried by quarks, almost
until CQHt is (almost) complete. This would mean that the residual plasma
component would gradually become richer in $B$, complicating the last
stages of CQHt, when bulk viscosity could reappear and even the option
of a first order transition might reopen.

In this work we shall not deepen these speculations, also because no
significant experimental or lattice data can be exploited in such
analysis. Accordingly, systems will be assumed to evolve through
thermodynamical equilibrium states and no viscosity or heat conduction
will be taken into account.

The plan of the paper is as follows: In the next Section we shall
briefly outline some relations obtained from thermodynamical
considerations in a cosmological context. Section 3 will just be
devoted to show the lattice data we shall be using. In Section 4 we
shall fit such data with a suitable analytical expression, able to
reconnect them with the asymptotic freedom regime. In Section 5.
instead, we shall focus on the low--$T$ region and discuss how to
apply a Hagedorn--like expression to work out energy density and
pressure.  Then, in Section 6, the use of interpolating expressions,
in the intermediate range, will be discussed.  Cosmological
considerations begin from Section 7, where lattice outputs are
actually used to work out the cosmic expansion regime.  For the sake
of comparison, In Section 8 we also deal with a first order
transition, approaching lattice data. All that will enable us to plot
the connections between $a$ (the scale factor), $t$ (time) and $T$
(temperature); suitable combinations of such parameters and their
evolution will be shown in Section 9. In Section 10 we shall gine some
examples on the impact that the above detailed relations could have on
cosmological observables, in the era of precision cosmology.  Finally,
Section 11 is devoted to drawing our conclusions.

\section{Thermodynamics and cosmology}

If we let apart the $3q \to B$ reaction, there can be little doubts
that, around CQHt, relaxation times of particle interactions,
including neutrinos, are well below the cosmic time. Similarly, the
value of the baryon/entropy ratio, receiving no substancial
contribution after CQHt, allows to neglect the chemical potential
associated to $B$.

Within such context, starting from the thermodynamical identity $dU =
-p\, dV + T\, dS$ and setting $\epsilon = \partial U/\partial V$, $\sigma
= \partial S/\partial V$, we have that
\begin{equation}
\sigma\, T = \epsilon + p~.
\label{s1}
\end{equation}
Let us then consider the free--energy $F = U-TS$, such that $dF = -p\,
dV - S\, dT$. Comparing its second derivatives in respect to $V$ and
$T$ when taken in different order, one has that
\begin{equation}
\sigma = \partial p/\partial T
\label{s2}
\end{equation}
so that the dependence on $T$ of
\begin{equation}
\epsilon = -p + T \partial p/\partial T
\label{pepsilon}
\end{equation}
is fully determined, once $p(T)$ is known. If we then define
\begin{equation}
\Phi = p/T^4~,~~~ E = \epsilon/T^4~,
\label{PhiE}
\end{equation}
the above equation yields
\begin{equation}
E(T) = 3 \Phi(T) + T {\partial \Phi \over \partial T}~.
\label{PhiE1}
\end{equation}
If $E$ and $\Phi$ depend only on $T$, this equation can be seen also
as a differential equation with $\Phi(T)$ unknown. It is then
equivalent to the relation $T^2 E = d(T^3 \phi)/dT$ and can be soon
integrated yielding
\begin{equation}
\Phi(T) = { T_r^3 \over T^3} \Phi(T_r) + {1 \over T^3} \int_{T_r}^T
d\tau\, \tau^2 E(\tau)~,
\label{integ}
\end{equation}
$T_r$ being a reference temperature. During CQHt, when $\Phi$ and $E$
actually depend just on $T$, eqs.~(\ref{PhiE1}) and (\ref{integ})
can be used to work out $\epsilon$ from $p$ and {\it viceversa}.

Should the transition be first order, however, we have two phases.
The quark--gluon plasma would have a pressure $p_{qg}(T)$, while the
hadron gas has a pressure $p_{h}(T)$ and the critical temperature
$T_c$ is when $p_{qg}(T_c) = p_{h}(T_c)$. It must however also be $
p'_{qg}(T_c) > p'_{h}(T_c)$ (here ' indicates differentiation in
respect to $T$) and therefore, according to eq.~(\ref{pepsilon}),
$\epsilon_{qg}(T_c) > \epsilon_{h}(T_c)$. 

When $T_c $ is reached, therefore, hadron bubbles must nucleate inside
the plasma. If this requires no substantial supercooling, the Universe
then undergoes an expansion at constant temperature, while the
fraction of each horizon occupied by the plasma decreases and the
fraction occupied by hadrons increases. Over large scales, the cosmic
energy density then reads
\begin{equation}
\epsilon = \epsilon_{qg} (1-\lambda) + \epsilon_h \lambda + 
3 \Phi_{l\gamma} T_c^4~.
\label{epsvar}
\end{equation}
Here $\lambda$ is the fraction of space occupied by the hadron gas,
while the pressure of the cosmic lepton--photon component, obtainable
from
\begin{equation}
\Phi_{l\gamma} = {\pi^2 \over 90} \left[{\rm N}_{bos} + {7 \over 8}
{\rm N}_{fer} \right] \simeq 1.5627~,
\label{lgamma}
\end{equation}
depends on the numbers ${\rm N}_{bos,fer}$ of the boson, fermion
relativistic spin states in the thermal soup. The above value arises
from assuming 14.25 effective spin degrees of freedom, {\it i.e.} that
$\mu$ particles are still fully relativistic. In Appendix A we show
that, in the $T$ range considered, this is acceptable. When aiming at
precise quantitative outputs, however, the fact that we are in a
region where $m_\mu \sim T$ should not be disregarded.

If supercooling is quite small or infinitesimal, a first order phase
transition implies quite a small or infinitesimal entropy input.  In
fact, let $S = a^3 \sigma = (a^3/T) (\epsilon + p)$ be the comoving
entropy. If $T = T_c - \delta T$ (with infinitesimal $\delta T$), the
average cosmic pressure will be
$$
p ~=~ p_{qg} (1-\lambda) + p_h \lambda + \Phi_{l\gamma} T_c^4  \simeq
~~~~~~~~~~~~~
$$
\begin{equation}
\simeq p(T_c) -
[\epsilon_{qg}(T_c) (1 - \lambda) + \epsilon_{h} (T_c) \lambda] {\delta
T \over T_c }
\label{pvar}
\end{equation}
so that, if $\delta T$ does not depend on $\lambda$,
\begin{equation}
{\partial p \over \partial \lambda} = 
[\epsilon_{qg}(T_c) - \epsilon_{h} (T_c) ] {\delta
T \over T_c }.
\label{pavar1}
\end{equation}
Let us consider then the cosmological context, where the space--time
metric reads
\begin{equation}
ds^2 = c^2 dt^2 - a^2(t) d\ell^2~,
\end{equation}
$d\ell$ being the infinitesimal comoving length element. Owing to
the Friedman equation
\begin{equation}
d[a^3 (\epsilon + p)] = a^3 dp
\label{fried1}
\end{equation}
it is then
\begin{equation}
\dot S =  (a^3/T) \dot p - S  (\dot T / T) =
\nonumber
\end{equation}
\begin{equation}
~~~= (a^3/T)\, (\partial p/\partial \lambda)\, \dot \lambda
+ a^3 (\partial p/\partial T)\, (\dot T/T) -
S (\dot T / T) 
\label{S1}
\end{equation}
(dots indicate here time differentiation) and the last two term at the
r.h.s. cancel, owing to eq.~(\ref{s2}), so that
\begin{equation}
\dot S = a^3 {\epsilon_{qg}(T_c) - \epsilon_{h} (T_c) \over T_c} \,
{\delta T \over T_c }\, \dot \lambda  = [S_{qg}(T_c) - S_{h}(T_c)]
{\delta T \over T_c } \dot \lambda
\end{equation}
fully vanishes as $\delta T/T_c \to 0~.$ This equation shows that only
a tiny fraction of the entropy difference between the two phases may
enrich cosmic entropy. The rest of the difference turns into
lepton--photon entropy. This is why neglecting the latter component
is strongly misleading.

When dealing with a {\it supposed} first order transition, we shall
then consider a negligible supercooling and $S$ to be constant. Any
other choice would require further assumptions.

\section{Lattice QCD data}

Lattice QCD deals with the non--perturbative regime for the QCD
equation of state. Here we are referring to recent results obtained
from \cite{cheng}. They used a $(N_t=6) \times 32^3$ size lattice to
calculate the pressure $p$ and the {\it trace anomaly} $\epsilon -3p$
for 20 values $T_i$ of temperature. The computation is performed using
physical values for the masses of the {\it up}, {\it down} and {\it
strange} quarks. No error is given for $p_i = p(T_i)$ estimates.  The
trace anomaly, instead, has a significant uncertainty which therefore
affects their $\epsilon_i = \epsilon(T_i)$ estimates.

These estimates are reported in the Table herebelow ($T$'s in units of
100~MeV, the other quantities in units of $(100~{\rm MeV})^4$).

From this table the estimates of the coefficients $E(T)$ and $\Phi(T)$
can be deduced; they will be shown in a number of Figures hereafter.

\vskip .5truecm

$$
\matrix{
 &	{\bf Table} &	{\bf I} \hfill &	 \cr
{\bf } &	 &	 &	 \cr
T  & \epsilon & p & \Delta \epsilon \cr
{\bf } &&& \cr
7.19 &	39300 &	12600 &	353 \cr
5.56 &	13800 &	4300 &	164.5 \cr
4.68 &	6800 &	2070 &	95.2 \cr
4.19 &	4350 &	1300 &	65.3 \cr
3.67 &	2520 &	714 &	48.02 \cr
3.24 &	1520 &	398 &	37.92 \cr
2.82 &	859 &	196 &	10.54 \cr
2.61 &	635 &	127 &	10.82 \cr
2.40 &	449 &	74.1 &	6.63 \cr
2.27 &	341 &	47.3 &	7.08 \cr
2.13 &	238 &	26.6 &	5.49 \cr
2.06 &	196 &	18.6 &	6.3 \cr
2.03 &	170 &	15.3 &	5.66 \cr
1.96 &	113 &	10.1 &	4.16 \cr
1.86 &	58.7 &	5.35 &	2.81 \cr
1.80 &	38.7 &	3.57 &	1.85 \cr
1.74 &	21.4 &	2.45 &	2.15 \cr
1.59 &	8.88 &	.909 &	1.2 \cr
1.45 &	6.24 &	.315 &	1.15 \cr
1.40 &	2.54 &	.189 &	0.9 \cr
}
$$
\section{The high--$T$ regime}
At high $T$, we expect $E = \epsilon/T^4~$ to approach asymptotically
a value
\begin{equation}
\tilde E = 3\, \tilde \Phi = (\pi^2/30) [16+(7/2)N_{col}N_{fla}]
\label{asfr}
\end{equation}
$N_{col}$ and $N_{fla}$ being the number of colors and flavors.  The
above estimates of $\epsilon $ and $p$ show that, at $T = 719\, $MeV,
it is $E \simeq 0.94 \times \tilde E~~ (= 15.627$, with $N_{col}=3$
and $N_{fla}=3$). In order to study the CQHt, however, we need an
expression telling us how $E(T)$ and $\Phi(T)$ approach $\tilde E$ and
$\tilde \Phi$ above 719 MeV.

The expression we shall propose, however, extends its validity down to
$T \sim 200\, $MeV, fitting lattice outputs all through this $T$
interval, being then suitable to deal also with a wide $T$--range
where lattice outputs are available, and not only because it provides
an easily treatable fit to them.

Let us then assume
\begin{equation}
\Phi = \tilde \Phi \left\{ 1 - \left[{T_o  \over T } (1 + \delta_o)
\right]^{s_o} \right\}
\label{e1}
\end{equation}
with
\begin{equation}
\delta_0 = \left[{T_1 \over T }(1 + \delta_{1}) \right]^{s_{1}},~~
\delta_1 = \left[{T_{2} \over T } (1 + \delta_{2})
\right]^{s_{2}},~ ......
\label{e2}
\end{equation}
an expression which, {\it a priori}, depends on the successions $\{
T_i \}$, $\{ s_i \}$. 

By using eq.~(\ref{PhiE1}), it is then possible to work out also the
recursive expressions:
\begin{equation}
E = 3\Phi + s_0(\tilde \Phi-\Phi)\left(1 - {T \dot \delta_0 \over
1 + \delta_0} \right)
\end{equation}
with
\begin{equation}
T \dot \delta_r = -s_{r+1} \left(1 - {T \dot \delta_{r+1} \over
1 + \delta_{r+1}} \right) \delta_r
~~~~~ (r=0,1,....)
\end{equation}
These expressions can be truncated at any order $\bar n$, by assuming
$\delta_{\bar n} \equiv 0$.

\begin{figure}
\resizebox{0.52\textwidth}{!}{%
\includegraphics{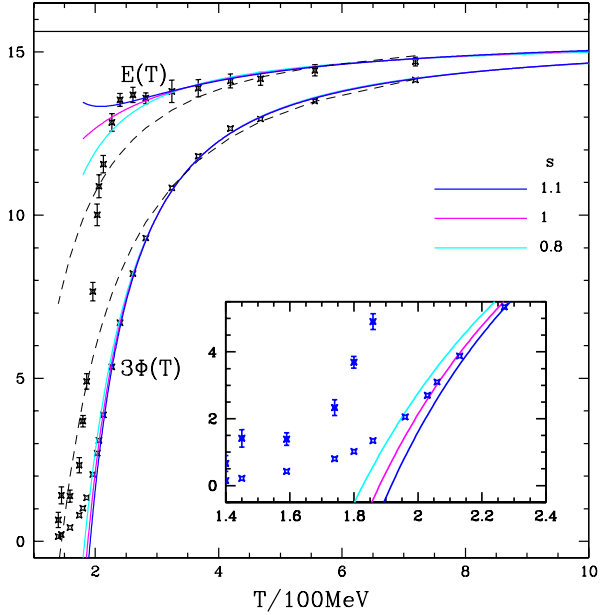}
}
\caption{\sl Lattice outputs compared with the high--$T$ fitting
expressions. The $E(T)$ points are reported with their 1--$\sigma$
errors. The $\Phi(T)$ points are reported without errors (not provided
in the lattice outputs used).  Dashed curves are obtained for
$\delta_0 \equiv 0$.  Continuous curves, instead, are obtaned for
$\delta_0 \neq 0$ and $\delta_1 = 0$. They are fits obtained with the
procedure described in Appendix B, however freely selecting the $s_o$
parameter.  In the case considered, the expressions (\ref{e1}),
(\ref{e2}) allow to approximate up to 10 (over 20) $E$ lattice values
(and $\sim 14$ $\Phi$ points). They are no longer suitable when the curvature
of $\Phi(T)$ changes direction.  In the inner frame, the low--$T$
behavior of $\Phi(T)$ is magnified.  The upper black line yields
$\tilde E$, {\it i.e.} $E(T) = 3\Phi(T)$ for an ideal gas of
relativistic quarks and gluons. }
\label{hTmake3}
\vskip .05truecm
\end{figure}

If we take $\delta_o = 0$ and $s_o=4$ we recover the MIT Bag--model
expression with a bag constant $B = \tilde \Phi T_o^4$. Such
expression is known not to fit lattice outputs; however, the values
$s_o \sim 1.48$ and $T_o = 145\, $MeV, are the best--fit to the
highest 6 $\Phi$ points. The quality of this fit is shown in Figure
\ref{hTmake3}.

A much better fit is however obtainable if we keep $\delta_o \neq 0$
and set $\delta_1 = 0$. In Figure \ref{hTmake3} we show some results
obtained by freely selecting $s_0$ and working then out $s_1$, $T_o$
and $T_1$ from the highest 6 $\Phi$ points, as above. The detailed
procedure followed in described in Appendix B.

Clearly, the presence of the $\delta_1$ term has the result of
increasing the bending in the middle--$T$ area: the expression
(\ref{e1}) sets a {\it $T$--point} where the pressure turns
negative. (We do not expect such point to be reached. The hadron gas
should replace the quark--gluon plasma before then.) Adding a
$\delta_0 ~~(> 0)$ correction means that, while $T$ decreases (and the
quark--quark distance increases), the {\it $T$--point} gradually shifts
to a higher value. 

One could argue that subtracting from the ideal gas coefficient the
power law term $(T_o/T)^{s_o}$ is a reasonable way to model the
binding action due to gluon exchange, when the cosmic expansion tries
to separe quarks beyond their confinement distance. The swiching on of
the second power law correction $\delta_o = (T_1/T)^{s_1}$ could then
be reminiscent of the expectation that, as the inter--quark distance
increases furthermore, the exchanged confining gluons, being colored,
may become source of further gluon emission, so causing a further
strengthening of their binding action.

From Figure \ref{hTmake3} we also see that reasonable $s_0$ values are
between 1 and 1.1~. Already at $s_0= 1.1$ the $E(T)$ dependence
exhibits a rebounce, which becomes more pronounced and shifts to
higher $T$'s for greater $s_0$ values; $s_0 = 1.1$ appears then as an
optimal choice to meet nine $E(T)$ error bars (for $T_i$ with $i =
12,....~20$), with no rebounce above $T_{12}$. If one takes a smaller
$s_0$ ({\it e.g.}, $s_0 = 1$), it is possible to meet also the
$T_{11}$ error bar, at the expences of shifting below $T_{12}$ and
$T_{13}$. (Figure \ref{hTmake3} apparently indicates that the fit with
$\Phi(T_i)$ values is fair, even down to $i \sim 6$--7, well below
12. In the inner box we however magnify the low--$T$ area and show
that, although the absolute values of $\Phi(T_i)$ are really well
approached, the expression (\ref{e1}) misses the shift from positive
to negative second derivative, essential to avoid the rebounce of
$E(T)$ towards low $T$'s.)

As a matter of fact, a reliable analytical expression, meeting half
lattice points and connecting the range of lattice data with
asymptotic freedom, is to be implemented by a numerical fit to lattice
data at lower $T$'s. It seems clear that the expressions (\ref{e1})
and (\ref{e2}), with $\delta_1 = 0$ and $s_0 \simeq 1.1$ are suitable
to this aim.

This choice is somehow corroborated by some {\it curious} numerical
regularities. When $s_0 = 1.1$ is chosen, the data fit returns $s_1
\simeq 2\, s_0 = 2.2$, $T_o = 76.56\, $MeV and $T_1 = 229.7\, $MeV; so
that the $T_1/T_o$ ratio differs from 3 by $\sim 2:10^{4}$. Such
regularities hold in a small interval around 1.1 (typically between
$\sim 1.08$ and 1.12).

The expressions with $s_0 = 0.8$ will also be used, when aiming to
mimic a first--order phase transition. In this ca\-se, the fitting
procedure yields: $s_1= 2.00$, $T_0 = 26.1~$MeV and $T_1 \sim 447\,
$MeV.

We also tried to add an extra term to eqs.~(\ref{e1})--(\ref{e2}),
allowing for $\delta_1 \neq 0$. The fitting procedure becomes then
even more intricate. A fit with lattice data is obtained with the
regular values $T_0=66\, $MeV, $T_1=2.15\, T_0$, $T_2=3.3\, T_0$,
while $s_0 = 1$, $s_1 = 1.5$, $s_2 = 2$. This fit, however, is no
improvement in respect to the one with $\delta_1 = 0\, $: it actually
meets the $E(T)$ error bars more at their centers, but has a rebounce
at greater $T$, making it useless already for $i = 12$; we avoid
further graphics complications and we refrain from showing it in
Figure \ref{hTmake3}.

\section{The low--T regime}
Expressions of $E(T)$ and $\Phi(T)$, for $T \sim 100$--$200\, $MeV may
hardly keep any reference to an ideal relativistic gas. As a matter of
fact, the lightest hadrons are pions and, if we tentatively assume for
them relativistic ideal gas expressions when $T \sim m_\pi$, we work
out that their mutual distance $D$, in average, is given by
\begin{equation}
D^{-3} \simeq {\zeta (3) \over \pi^2} 3 T^3 \simeq 0.37\, T^3
\end{equation}
($\zeta (3) \simeq 1.202$); this yields $D \sim 1.4/m_\pi$, a distance
comparable with hadron hard core. Accordingly, hadron proper volumes
cannot be neglected, let alone hadron--hadron interactions.
Furthermore, as $T$ increases, we can hardly neglect heavier hadronic
resonances.

There have been several approaches to the thermodynamics of the
hadronic gas. In any case, however, because of their volume and of the
temperature, hadron will never yield a major cosmic component.
Leptons and photons will then have a density coefficient $E_{l\gamma}
\sim (\pi^2/30)15 $, while hadrons $E_s \ll (\pi^2/30)
3$. Accordingly, for our aims, taking the expressions inspired to the
Hagedorn model will be fully reasonable.  Such expressions contain 3
constant, that we shall tune to the low--$T$ lattice outputs.

The pressure of a Hagedorn gas with vanishing chemical potential reads
then
\begin{equation}
\label{h1}
p(T) = {1 \over 6 \pi^2} \int_{m_\pi}^{T_H} dm\, w(m) \int_0^\infty
dk\, {k^4 E^{-1}(k,m) \over \exp[E(k,m)/T]-1 }
\end{equation}
with $E(k,m) = \sqrt{k^2 + m^2}$ and
\begin{equation}
\label{h2}
w(m) = \left(\bar m \over m \right)^\gamma \exp \left(m \over T_H \right)~.
\end{equation}
Selecting $\gamma = 5/2$ allows a closed expression, using just the
exponential integral function
$$
Ei(x) = \int_x^\infty {dt \over t} e^{-t}~,
$$
which reads
$$
p = \Phi(T) T^4 ~~~{\rm with} ~~~ ~~~~~~~~~~~~~~~~~~~~~~~~~~~~~~~~~~
$$
\begin{equation}
\label{phihag}
\Phi(T) = \alpha_0 
\left( T / T_{H} \right)^{5/2} Ei \left[m_0 (T^{-1}-T_H^{-1})
\right]
\end{equation}
and depends on the Hagedorn temperature $ T_H \simeq 200\, $MeV and on
the constants $\alpha_0$ and $m_0$, related to $\bar m$.

\begin{figure}
\resizebox{0.45\textwidth}{!}{%
\includegraphics{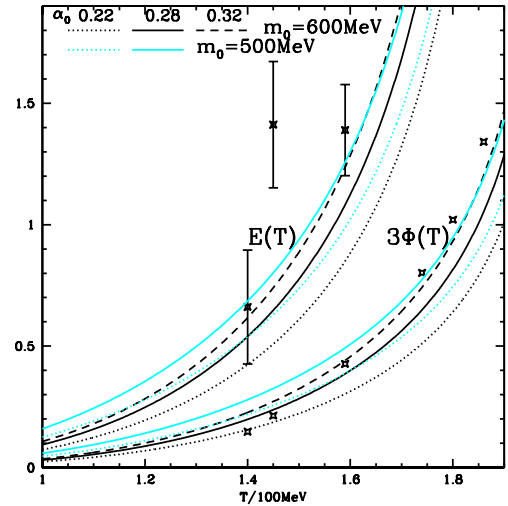}
}
\caption{\sl Lattice outputs compared with the low--$T$ Hagedorn--like
expressions. The $E(T)$ points are reported with their 1--$\sigma$
errors. The $\Phi(T)$ points are reported without errors (not provided
in the lattice outputs used). Although the expressions meet the trend
and the order of magnitude of lattice outputs, the high level of the
secont $E(T)$ point, if real, is clearly missed. }
\label{hag}
\vskip .05truecm
\end{figure}

From this expression of $\Phi(T)$ and eq.~(5) one easily works out that
\begin{equation}
E(T) = { 3 \over 2} \Phi(T) + \alpha_0 \left(T_H \over T \right)^{5/2}
{\exp \left[m_0 (T^{-1}-T_H^{-1}) \right] \over T_H/T - 1 }~;
\end{equation}
Figure \ref{hag} then shows the $\Phi(T)$ and $E(T)$ behaviors at
low--$T$, for 5 reasonable choices of the $\alpha_o$ and $m_0$
constants.
Although all parameter choices meet the general trend and the order of
magnitude of lattice outputs, there is a specific feature that none of
them approaches, the second $E(T)$ point, which is however high.

Accordingly, we shall use such expressions, in association with
lattice outputs, just for $T < T_1$~. Almost arbitrarily we chose then
$\alpha_0 = 0.28$, $m_0 = 600\, $MeV. The same values will also be
used when trying to mimic a first order transition, up to the
temperature where the expression (\ref{phihag}) of $\Phi(T)$ crosses
the high--$T$ expression (\ref{e1}).

\begin{figure}
\resizebox{0.50\textwidth}{!}{%
\includegraphics{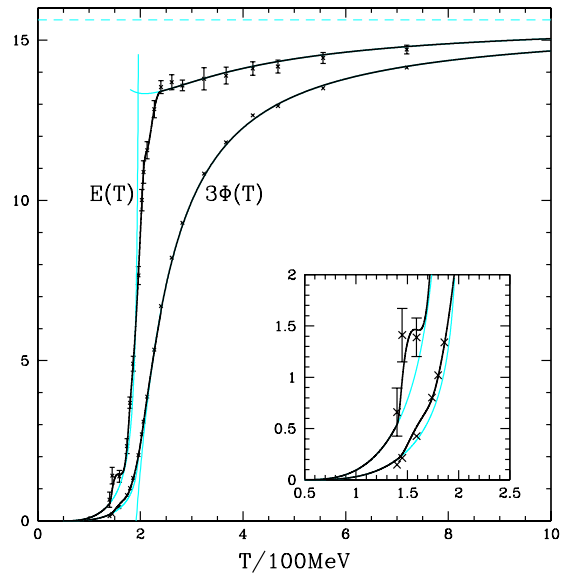}
}
\caption{\sl Lattice outputs and their interpolation, as described in
the text. The cyan solid lines represent the behavior of low--$T$ and
high--$T$ extrapolating expressions, out of their validity range. }
\label{npt}
\vskip -.05truecm
\end{figure}
\section{Fitting lattice outputs}
In order to use lattice outputs in Friedmann equations, one needs
interpolating them. In principle, mathematical and numerical libraries
contain excellent interpolating routines, namely the {\it splint} and
{\it spline} programs in {\it Numerical Recipes} \cite{NR}, making use
of a cubic polynomial.

However, if one uses such recipe to interpolate $\Phi(T)$ (let us also
outline that interpolating $p(T)$ yields essentially identical
results, as a counterproof of the outstanding efficiency of available
routines), the critical difficulty arises when $E(T)$ is seeken.  The
passage from $\Phi$ to $E$ implies differentiation, and this blows up
small irregularities; the resulting $E(T)$ comprises several
unmotivated oscillations, of clearly numerical origin.

In particular, there are fair reason to believe that $E(T) $ should be
always increasing, as also lattice outputs indicate, towards its
asymptotic freedom value. On the contrary, interpolation yields $T$
intervals, between contiguous $T_i$, where $E(T)$ decreases.  This
effect is particularly evident for $i \simeq 1,~2,~3,...$ and at $i >
11$--12.

We eliminate this problem with 3 actions: (i) At high $T$ we simply
use the analytical expressions (\ref{e1})--(\ref{e2}), starting from
$i = 12$. (ii) We interpolate the central points of $E(T)$, instead of
$\Phi(T)$, and make use of the integral expression (\ref{integ}) to
obtain $\Phi(T)$. This will be done using $T_1$ as $T_r$, and taking
the values of $E(T_1)$ and $\Phi(T_1)$ given by Hagedorn--like
expressions with our selected parameters. (iii) We select the $E(T_i)$
values at $i=2$ and 3, at hand within a 2--$\sigma$ error bar, at the
minimal distance from the central points able to prevent the spline to
yield any $E(T)$ decrease. Clearly other options are possible, but
they appear more intricate.

The overall behaviors of $E(T)$ and $\Phi(T)$ is then shown in Figure
\ref{npt}, where the critical low--$T$ region is also magnified.

\section{Use of lattice outputs in Friedmann equations}
The $T$ dependence described in Figure \ref{npt} is then
used to integrate the Friedmann equations.

It is then convenient to consider first entropy conservation, yielding
that
\begin{equation}
[E(T) + \Phi(T) + 4\, \Phi_{l\gamma}] (a\, T)^3 =
4 (\tilde \Phi +  \Phi_{l\gamma}) (a_i T_i)^3~.
\label{Scons}
\end{equation}
Here $a_i$, $T_i$ are scale factor and temperature at the ``initial''
conditions, where asymptotic freedom holds. In the computations here
we took $T_i = 80$ MeV, so to avoid intereferences with the
electroweak transition. From eq.~(\ref{Scons}) it is easy to work out
the $T(a) $ dependence and the deviations of the $a\, T$ product from
constant.

The dynamical Freedman equation then yields
\begin{equation}
\left( \dot a \over a \right)^2 = H_i^2 {E[T(a)] +  3 \Phi_{l\gamma}
\over 3( \tilde \Phi + \Phi_{l\gamma})} \left[T(a) \over  T_i \right]^4~;
\label{fr1}
\end{equation}
by taking into account that the initial Hubble parameter $H_i \simeq
1/2t_i$ (in the radiative expansion regime), it is then
\begin{equation}
2\, \int_{a_i}^a
{d{\rm a} \over {\rm a}} \left( \tilde \Phi + \Phi_{l\gamma} 
\over E[T({\rm a})]/3 +
\Phi_{l\gamma} \right )^{1/2} \left[T_i \over T({\rm a}) \right]^2 = 
{t \over t_i} - 1
\label{fr2}
\end{equation}
In this way one obtains $a(t)$ and thence $T(t)$.

Using this technique we work out the behaviors shown in Figures
\ref{aTa}--\ref{tT2}. Before discussing them, let us however outline
how we proceed to build analogous curves holding in the case of a
phase transition.

\section{Strongly interacting matter in a first--order phase transition}
\begin{figure}
\resizebox{0.50\textwidth}{!}{%
\includegraphics{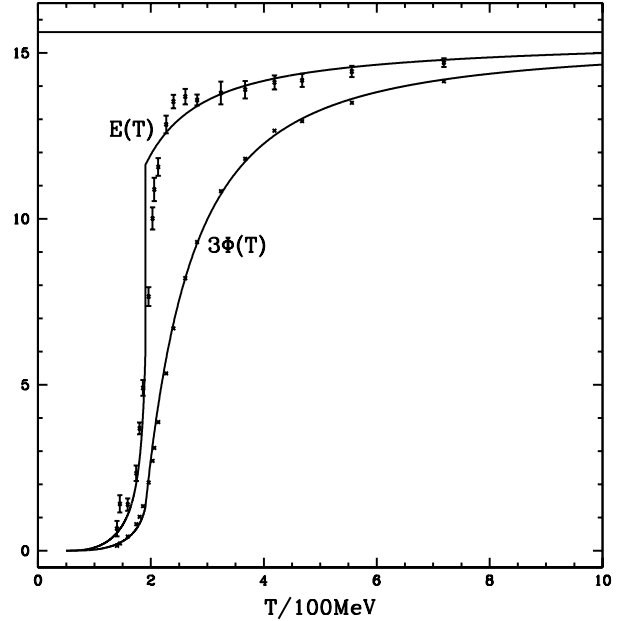}
}
\vskip -.2truecm
\caption{\sl Lattice outputs approximated by a first order phase
transition, as described in the text. Notice the vertical line in the
$E(T)$ curve, where the transition occurs. }
\label{ptshow}
\vskip -.05truecm
\end{figure}

In Figure \ref{ptshow} we show the $\Phi(T)$ curves we shall use to
mimic a phase transition: at high--$T$ we shall assume that $\Phi(T) $
is given by an expression (\ref{e1}) with $\delta_1 = 0$ and $s_0 =
0.8$; at low--$T$, but also at $T > T_1$, we shall use the
Hagedorn--like expression (\ref{phihag}) with the same parameters as
above. The two curves are taken above and below the temperature $T_c$
where they cross, respectively. Therefore, the overall $\Phi(T)$
behavior obtained in this way, although continuous, exhibits a
(modest) shift of slope at $T_c$. Accordingly, the low-- and high--$T$
curves for $E(T)$ do not intersect. In the Figure they are connected
by a vertical line.

A cosmological phase transition, occurring close to the critical
temperature, requires that, after $T_c$ is reached at a time $t_h$,
the Universe stops cooling down. In any volume an increasing fraction
of space will be then occupied by the hadron gas. When the high--$T$
plasma has completely vanished, at a time $t_l$, the cosmic
temperature restarts decreasing.

During all this process $S$ is conserved, according to eq.~(15).

Of course, one might also consider a transition occurring after a
significant supercooling. In order to work out the cosmic expansion
law, it would be necessary a supplement of information, yielding
{\it e.g.}, $\delta T/T$ as a function of $\lambda$.

The integration of Friedmann eqs. during a first order phase
transition occurring without supercooling was first performed in
\cite{bola}. Such integral was then rediscovered a few times by
various authors.

Let $p_c$ be the pressure at $T_c$, including also the lepton--photon
pressure. Also $\epsilon$ and $\sigma$ will include the lepton--photon
contribution, all through this Section. They will be labeled with $_h$
($_l$), to refer to the time when the transition begins (ends) and the
same labels will be used for time and scale factor. Let us also remind
that $\lambda$ is the portion of space occupied by the low--$T$ phase.

From the definition of comoving entropy, we have that
\begin{equation}
\label{Scons1}
\epsilon = T_c S /a^3 - p_c = \sigma_h T_c (a_h/a)^3-p_c
= \epsilon_h (1-\lambda) + \epsilon_l \lambda
\end{equation}
with $\sigma_h = S/a^3_h = (\epsilon_h + p_c)/T_c$. This eq.~sets a
link between $a$ and $\lambda$.

The dynamical Friedmann eq.~reads then
\begin{equation}
(\dot a/a)^2 = (8\pi G/3) [\sigma_h T_c (a_h/a)^3 - p_c]
\end{equation}
and, by taking $u^2 = (a/a_h)^{3}/(\epsilon_h/p_c+1),$ can be set into
the form
\begin{equation}
du~(1-u^2)^{-1/2} = (6\pi G p_c)^{1/2} dt
\end{equation}
whose integral reads
\begin{equation}
u = \sin[(6\pi G p_c)^{1/2} t + A]~,
\end{equation}
$A$ being determined by the conditions at $t_h$.

In the case of the $T$--dependence of $E$ and $\Phi$ shown in Figure
\ref{ptshow}, we shall use the same equations as in the absence of a
phase transition for $t < t_h$ and $> t_l$, and the above expressions
during the transition.

\section{Scale factor, temperature, time connections}
When the cosmic temperature shifts from some thousands MeV to $\sim
100 MeV$ and the CQHt occurs, scale factor and time increase by orders
of magnitude.

In order to describe the cosmic evolution, it is then worth showing the
behavior of precise combinations of $a$, $T$ and $t$, starting from
the {\it initial} values $a_{i}$, $T_i$ and $t_{i}$.

Let us then recall first that, in a radiation dominated expansion,
when the total number of spin states is constant, the product $aT$ is
also constant. Of course, during the CQHt, $aT$ will have robust
variations; entropy conservation prescribes that, at the end of the
transition, $ a_f T_f = a_i T_i (g_i / g_f)^{1/3} $ with $g_i/g_f
\simeq (47.5+14.25)/14.25$. Accordingly, the fact that $aT$ shall
increase by a factor $\sim 1.6$ can be soon predicted, and is
indipendent from the detail of the transition.

\begin{figure}
\resizebox{0.50\textwidth}{!}{%
\includegraphics{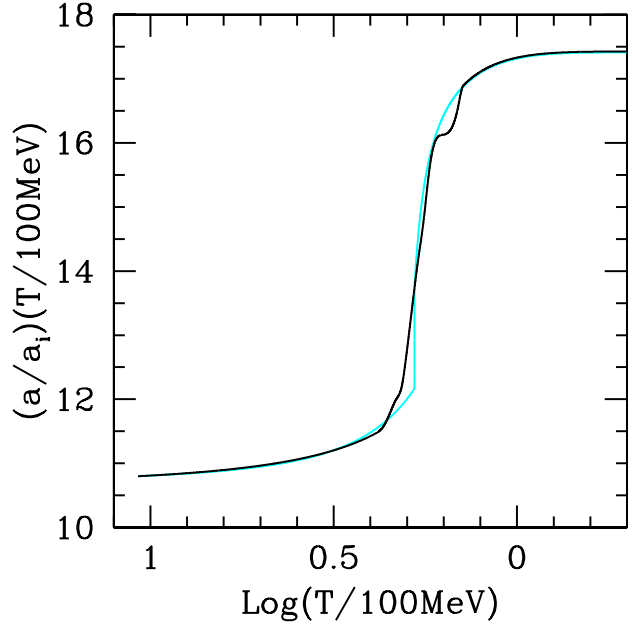}
}
\caption{\sl Dependence of the $aT$ product on the temperature. The
black curve is derived by using lattice data; the cyan curve would hold
in the case of a phase transition.  Notice the vertical increase of
$aT$ while $T$ remains constant during the phase transition. }
\label{Ta}
\end{figure}

\begin{figure}
\resizebox{0.50\textwidth}{!}{%
\includegraphics{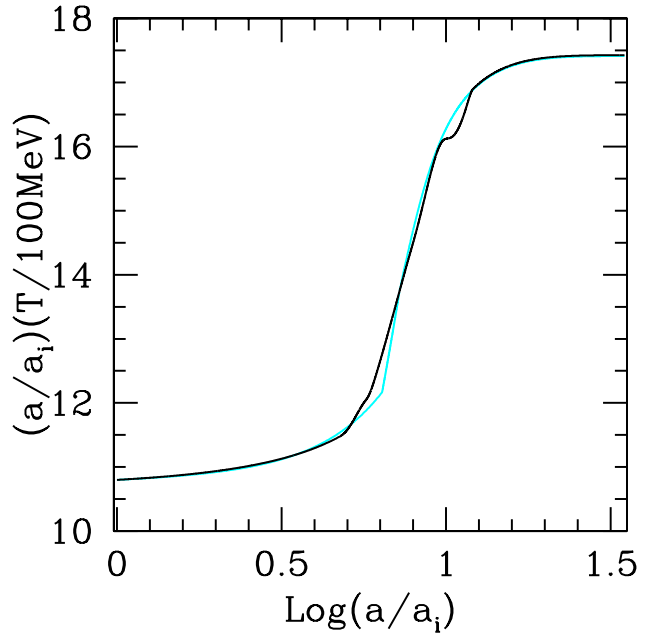}
}
\caption{\sl Dependence of the $aT$ product on scale factor. 
Black and cyan colors as in the previous Figure. }
\label{aT}
\end{figure}

Accordingly, in Figures \ref{Ta}--\ref{aT} we show how $aT$ varies, as a
function of the scale factor or the temperature, from its initial to
its final values, however known {\it a priori}.

Let us then notice, in particular, the feature close to the transition
end, when lattice outputs are used. It is due to the {\it anomaly} in
$E(T)$ at $T_2$ and similar features will be present in the next
plots. Such features cannot be certainly predicted from conservation
theorems.

\begin{figure}
\resizebox{0.50\textwidth}{!}{%
\includegraphics{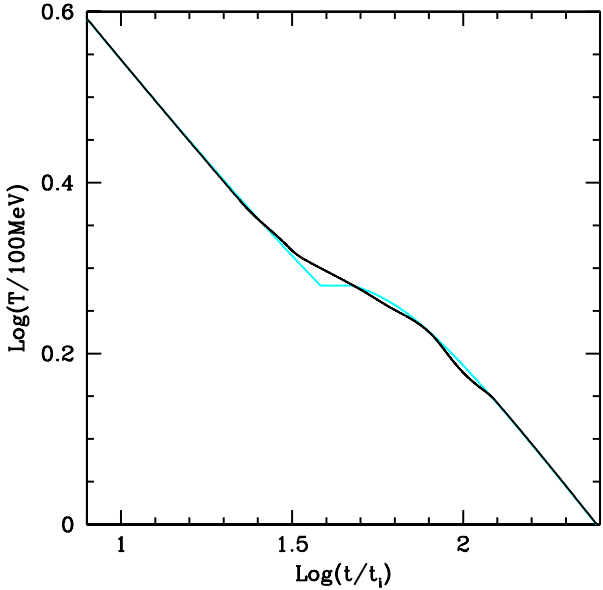}
}
\caption{\sl Time dependence of $T$. Black and cyan colors as in
previous Figures. Notice again the flat interval in the cyan curve:
during a first order phase transition $T$ stays constant while time
elapses. }
\label{Tot}
\end{figure}

In Figure \ref{Tot} we then exhibit the time dependence of $T$.
The discrepancies between the two evolutions are evident, during the
transition. On the contrary, a slight offset between the two curves,
after tre transition is completed, is just noticeable.

As a matter of fact, also in this case we must recall that, during a
radiative expansion when the number of spin degrees of freedom does
not vary, $tT^2$ should be constant. This derives from the dynamical
Friedmann equation, when it is written in the form
\begin{equation}
\label{tt2}
(1/2t)^2 = (8\pi^3 G/90) g\, T^4~,
\end{equation}
true if the relation $a \propto t^{1/2}$ is strictly valid.

Even though we assume eq.~(\ref{tt2}) to be true {\it before} the
transition, not only during the transition itself eq.~(\ref{tt2})
looses its validity, but also when $g$ would be stabilized at a {\it
final} value, we can only expect that $a \propto (t+\bar {\delta
t})^{1/2}$ with a suitable $\bar {\delta t}$ correction. This is what
originates the minimal offset of the final curves in Figure \ref{Tot}.

In principle, therefore, the time and scale dependence of $tT^2$ are
relevant both for the transition period and for the final settlement.

\begin{figure}
\resizebox{0.50\textwidth}{!}{%
\includegraphics{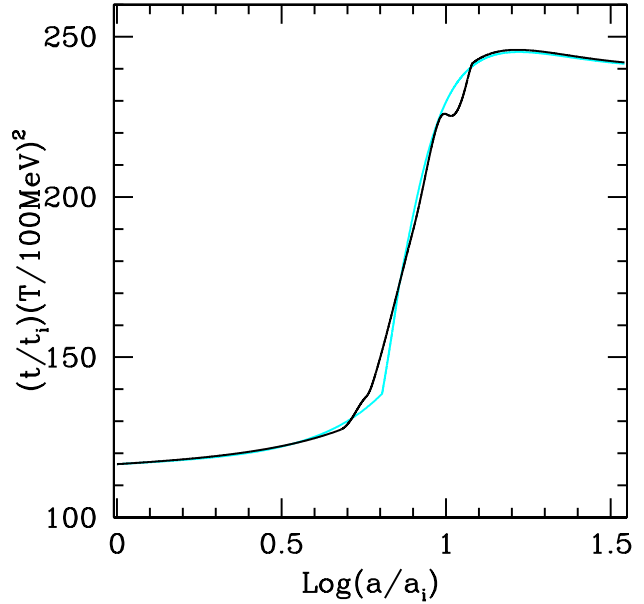}
}
\caption{\sl Time dependence of $tT^2$. Black and cyan colors as in
previous Figures. Notice again the feature arising from the $E(T_2)$
value. }
\label{tT2}
\end{figure}

In Figure \ref{tT2} we show then the explicit dependence of $tT^2$ on
the scale factor $a$. Here again the $E(T_2)$ anomaly causes a clear
feature.
But the offset at the end of the transition, once again, is just
barely noticeable.

This leads us to conclude that, even when we assume significantly
different $E(T)$ and $\Phi(T)$ behaviors, the relations among $T$, $a$
and $t$ exhibit discrepacies mostly during the transition.  We should
select completely awkward --~and unlikely~-- $E(T)$ and $\Phi(T)$ to
cause a substantial shift in $T(t)$ at the transition end.

We then conclude that, once the transition is completed, no
significant dependence on the relations among $T$, $a$ and $t$
remains: a regime trend is recovered and, if no other quantity was
altered {\it during} the transition, one can hardly expect signals
from the relative values of such parameters.

\section{Observable quantities}
As an example of possible observables, let us consider
non--relativistic Majorana spinors, mutually annihilating with a cross
section $\sigma_a$, such that the annihilation time $\tau = (n \langle
\sigma_a v_T \rangle )^{-1}$ is $ O$$(10^{-6}$--$10^{-5})~{\rm sec.}$,
occurring then across the quark--hadron transition. Here $n$ is the
particle number density, and $v_T$ is their mutual velocity, in the
temperature conditions considered.

Their comoving number density $n_c = na^3$ shall obey the equation
\begin{equation}
\label{nc}
dn_c/dt + n_c^2 \langle \sigma_a v_T \rangle/a^3 = 0
\end{equation}
and one must know the time dependences of $\langle \sigma_a v_T
\rangle$ and $a^3$, in order to integrate such equation. At large $t$,
such integral must yield $n = \bar n (a_{dg}/a)^3$, $a_{dg}$ being the
scale factor at a time $t_{dg}$, the {\it decoupling time}.  This
means that the $t \propto a^2$ proportionality is normalized so that
$t/t_{dg} = (a/a_{dg})^2$.

We shall however be cautious, before assuming that $t_{dg}$ coincides
with $\tau$, so that $\bar n = n(t_{dg})$. For instance, assuming a
purely radiative expansion during the decay stages and $\langle
\sigma_a v_T \rangle = {\rm const}$, one obtains that
\begin{equation}
\label{nt}
n(t) = {n(t_{dg})
\over
1 + 2(t_{dg}/\tau)
[1 - (t_{dg}/t)^{1/2}]} \left[a_{dg} \over a \right]^3~.
\end{equation}
Therefore, at $t \gg t_{dg}$, it shall be $n(t) = \bar n(a_{dg}/a)^3$
with
\begin{equation}
\bar n = n(t_{dg})/(1 + 2t_{dg}/\tau)~.
\end{equation}
Assuming $\bar n \equiv n(t_{dg})$ is therefore just a zero--th order
approximation; on the contrary, taking $\tau = t_{dg}$, we see that
the $\bar n/ n(t_{dg})$ ratio is $\simeq 0.33~.$ A different $T$
dependence of $\langle \sigma_a v_T \rangle $, or a different $t$
dependence of $a$, would leave most of this treatment unchanged, just
modifying the factor 2 in front of $(t_{dg}/\tau)$ in eq.~(\ref{nt}).
Accordingly, the detailed $t$ dependence of $a$ (or $\langle \sigma
v_T \rangle$) may change the residual amount of uninteracting Majorana
spinors, possibly constituting CDM, by a factor {\it O}$(2)$.

The point is that cosmology is becoming a precision science and a
variation by a fraction of percent, in the abundance of CDM, may soon
be measured. Accordingly, a percent approximation in the value of
$n_{dg}$, let alone an uncertainty by a factor {\it O}$(2)$, might
soon become appreciable.

If one aims then at connecting cosmological data with microphysical
data, such as $\sigma_a$ and its $T$ dependence, a precise knowledge
of $a(t) $ during the CQHt could be strictly required.

\section{Conclusions}
This paper was devoted to updating the treatment of the cosmological
Quark--Hadron transition, taking into account recent lattice outputs.

Such transition had seemed to bear a major relevance, in the
mid--Eighties, when most reserachers seriously considered the option
that it could be a first order phase transition. When lattice outputs
suggested that it was a crossover, its interest faded.

With the coming of the era of precision cosmology, however, a detailed
knowledge of expansion law and thermal history, when the time elapsed
since the Big--Bang was {\it O}$(10^{-6}$--$10^{-5}~{\rm sec.})$, could
become vital to interconnect cosmological observables and
microphysical parameters.

In this work we have seen that the most recent lattice data are still
hard to be used to work out expansion law and thermal history. This is
due, first of all, to the uncertainly on the $E(T)$ behavior, deduced
from the trace of the stress--energy tensor, still given with wide
error bars. Furthermore, even most recent data concern typically 20
temperature values $T_i$ and, perhaps, they are still not enough; we
should need more $T_i$ points.

For instance, at the low--$T$ end, when strongly interacting matter
can be considered a $\pi$+resonances gas, there is a clear anomaly in
the $E(T)$ trend, however outlined by a single point $E(T_2)$.  Such
anomaly, {\it e.g.}, vanifies the use of Hagedorn--like expressions in
this $T$--range. In our treatment, we took such anomaly into account
and saw that it causes evident features also in the cosmological
expansion law. We however legimately wonder whether such features,
relying on a single estimate among 20, are to be considered safe.

In the attempt to overcome a part of such difficulties, however, we
have proposed an expression able to fit almost half of the lattice
points in the high--$T$ range, and to connect them with the very high
$T$ regime. In our opinion this expression, which {\it a priori}
depends on 4 parameters, but nicely fits more than $ 20$ $\epsilon$
and $p$ values, could also allow some insight into the physics of the
plasma; as a matter of fact, the best fit of the above parameters led
to values unexpectedly regular, so that the expression might prove to
be something more than a fitting algorithm.

The main results of this paper are however well summarized in the
Figures and their captions, that can be useful to convey the overall
message of this work.

\vskip .3truecm
\noindent
{\bf Acknowledgment}. Thanks are due to Maria Paola Lombardi for
encouraging us to proceed in the preparation of this paper and for a
number of interesting discussions.

{}

\centerline{\bf Appendix A}

\begin{figure}
\resizebox{0.40\textwidth}{!}{%
\includegraphics{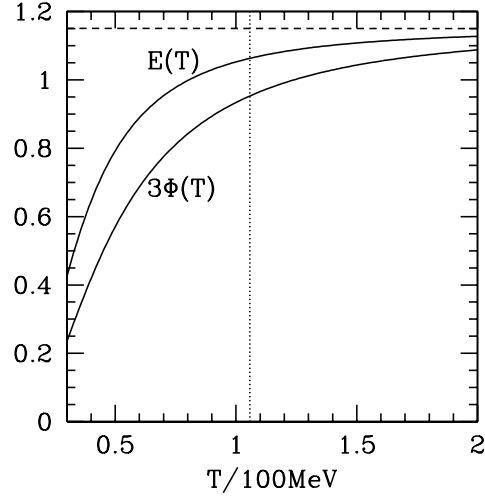}
}
\caption{\sl Temperature dependence of the energy density and pressure
of the muon thermal component. The vertical dotted line indicates the mass;
the dashed horizontal line are the asymptotic ultrarelativistic values.
Clearly, for $T \sim m_\mu$, energy density and pressure have already
$\sim 90\, \%$ of their asymptotic values. }
\label{muons}
\end{figure}

In order to evaluate the energy density and pressure due to $\mu$
particles, which are becoming non--relativistic in the $T$--range
considered, one must perform the integrals
$$
\epsilon_\mu (T) = {2 \over \pi^2} \int_0^\infty dk {k^2 W(k)
\over \exp[W(k)/T] + 1 }
~,$$$$
3p_\mu (T) = {2 \over \pi^2} \int_0^\infty dk {k^4  W^{-1}(k)
\over \exp[W(k)/T] + 1 }
\eqno (A1)
$$ 
where $W(k) = \sqrt{k^2 + m_\mu^2}$, $m_\mu$ being the mass of the
particles. If one sets then $\epsilon_\mu = ET^4$, $p_\mu = \Phi
T^4$, it is easy to see that
$$
E(T) = {2 \over \pi^2} \int_0^\infty dx~ {x^3 Z(m_\mu/Tx)}{ \exp(x)
\over \exp [xZ(m_\mu/Tx)] + 1}
~,
$$$$ 3\Phi(T) = {2 \over \pi^2} \int_0^\infty dx ~{x^3 Z^{-1}(m_\mu/Tx)
}{ \exp(x) \over \exp [xZ(m_\mu/Tx)] + 1 } ~
\eqno (A2)
$$
with $Z = \sqrt{1 + (m_\mu/Tx)^2}$. These integrals can be easily performed
numerically and yield the results shown in Figure \ref{muons}.

Accordingly, around 100~MeV, energy density and pressure of the muon
component have not yet substantially abandoned their ultrarelativistic
values.

\vskip .4truecm
\centerline{\bf Appendix B}

\vglue .5truecm

In order to fit the expression (\ref{e1}) with data, we follow
these steps:

We select a value for $s_o$ and use lattice outputs to obtain
$$
\psi_i =  T_i [1-\Phi(T_i)/\tilde \Phi]^{1/s_o}~.
\eqno (B1)
$$ 
We directly use just the six $\psi_i$ values for $i=15,...,20$.
According to the expression (\ref{e1}), we expect that, in the
$T_{15}$--$T_{20}$ interval, at least, it is
$$
\psi(T) = T_0 + \varphi T^{-s_1}
\eqno (B2)
$$ 
for suitable values of the parameters $T_0$, $s_1$ and $\varphi
~(=T_1^{s_1})$. This expression implies that
$$
\psi_i - \psi_j = \varphi (T_i^{-s_1} - T_j^{-s_1})
\eqno (B3)
$$ 
(with $i,j \in 15$--20) and
$$
{\psi_i - \psi_j \over \psi_k - \psi_r} =
{T_i^{-s_1} - T_j^{-s_1} \over
T_k^{-s_1} - T_r^{-s_1}} 
\eqno (B4)
$$ 
where, again, $i,j,k,r \in 15$--20. For any choice of $i,j,k,r$, it is
then possible to determine a value of $s_1$. There are however 15
possible combinations, yielding 15 values $s_1$, that we can
average. Otherwise, we can directly treat the 15 combinations $
(T_i^{-s_1} - T_j^{-s_1}) / (T_k^{-s_1} - T_r^{-s_1}) $ ($i,j,k,r \in
15$--20) as independent data, seeking the best fit $s_1$ value. For
reasonable $s_o$ values we expect the two procedures to yield the same
$s_1$; we found consistent outputs (within $1:10^5$) for $s_o$ between
0.8 and 1.25~.

Using such $s_1$, we also average among $\varphi$ values obtainable
from eq.~(B3), and then eq.~(B2) yields a set of $ T_o$ values which
are also suitably averaged.

Although using just the 6 top pressure values in this procedure, the
agreement with many more pressure and energy density points eventually
follows (as a matter of fact, only 4 points would be barely sufficient
if we exclude~$T_{17}$). 

\end{document}